# FastLMFI: An Efficient Approach for Local Maximal Patterns Propagation and Maximal Patterns Superset Checking

Shariq Bashir
National University of Computer and Emerging Sciences, FAST House, Rohtas Road, G-9/4, Islamabad, Pakistan
shariqadel@yahoo.com

A. Rauf Baig
National University of Computer and Emerging Sciences, FAST House, Rohtas Road, G-9/4, Islamabad, Pakistan
rauf.baig@nu.edu.pk

## Abstract

***Maximal frequent patterns superset checking plays an important role in the efficient mining of complete Maximal Frequent Itemsets (MFI) and maximal search space pruning. In this paper we present a new indexing approach, FastLMFI for local maximal frequent patterns (itemset) propagation and maximal patterns superset checking. Experimental results on different sparse and dense datasets show that our work is better than the previous well known progressive focusing technique. We have also integrated our superset checking approach with an existing state of the art maximal itemsets algorithm Mafia, and compare our results with current best maximal itemsets algorithms afopt-max and FP (zhu)-max. Our results outperform afopt-max and FP (zhu)-max on dense (chess and mushroom) datasets on almost all support thresholds, which shows the effectiveness of our approach.***

## 1. Introduction

Frequent itemset mining is one of the fundamental problems in data mining and has many applications such as association rule mining, inductive databases, and query expansion. Let *T* be the transactions of the database and *X* be the set of items ($X \subseteq \{1...n\}$). An itemset *X* is frequent if it contains at least $\sigma$ transactions, where $\sigma$ is the minimum support. An itemset *X* is maximal if it is not subset of any other known frequent itemset.

When the frequent patterns are long, mining all Frequent Itemsets (FI) is infeasible because of the exponential number of frequent itemsets. Thus algorithms for mining Frequent Closed Itemsets FCI [10, 12] are proposed, because FCI is enough to generate association rules. However FCI could also be exponentially large as the FI. As a result, researchers now turn to MFI. Given the set of MFI, it is easy to analyze many interesting properties of the dataset, such as the longest pattern, the overlap of the MFI, etc. MFI mining has two advantages over all FI mining. First, MFI mines small and useful rules, and second a single database scan can collect all FI, if we have MFI.

In our opinion efficient mining of MFI depends upon three factors. First, The mining approach used for FI determination: candidate-generate-and-test [1] or pattern growth [8]. Second, search space pruning techniques [4]. Third, MFI superset checking which takes *O(MFI)* time in worst case. In last 5 years lots of techniques have been developed for first two factors, but a very little consideration has been given to MFI superset checking. Zaki et al in [6] is one of the pioneers who showed the importance of superset checking. We also observed in our experiments that MFI superset checking cost is almost half of total MFI mining cost, especially for large sparse and dense datasets. This also shows the importance of this factor in overall MFI mining.

In this paper we propose a new indexing approach, FastLMFI for local maximal patterns propagation and maximal patterns superset checking, which is better than previous well known progressive focusing technique.

Rest of the paper is organized as follows; Section 3 presents a well known related work ***progressive focusing*** approach, Section 4 describes the structure of our FastLMFI (local maximal patterns propagation and maximal superset checking) approach. FastLMFI efficient implementation is presented in Section 5, while results are described in Section 6 and Section 7.

## 2. Preliminaries

Let < be some lexicographical order of the items in *TDB* such that for every two items *a* and *b*, $a \neq b$: $a < b$ or $a > b$. The search space of frequent itemset mining can be considered as a lexicographical order [15], where root node contains an empty itemset, and each lower level k contains all the k-itemsets. Each node of search space is composed of head and tail elements. Head denotes the itemset of node, and items of tail are possible extensions of new child itemsets. For example with four items {A, B, C, D}, at level 0 (root) head is empty 〈{}〉 and tail is composed with all of items 〈(A, B, C, D)〉, which generates four possible child nodes {head 〈(A)〉: tail 〈(BCD)〉}, {head 〈(B)〉: tail 〈(CD)〉}, {head 〈(C)〉: tail 〈(D)〉}, {head 〈(D)〉: tail 〈{}〉}. This MFI search space can be traversed by breadth first search or depth first search.

Let *list (MFI)* be our currently known maximal patterns, and let *Y* be our new candidate maximal pattern (itemset). To check if *Y* is subset of any known mined maximal pattern, we perform a maximal superset checking, which takes *O(MFI)* in worst case. To speedup the superset checking cost, local maximal frequent itemset (LMFI) has been proposed. LMFI is a divide and conquer strategy, which contains only those relevant maximal patterns, in which *Y* appears as a prefix.

Any maximal pattern containing *P* itemsets can be a superset of $P \cup_{subsets(P)}$ or $P \cup_{freq_ext(P)}$. The set of $P \cup_{freq_ext(P)}$ is called the local maximal frequent itemset with respect to *P*, denoted as $LMFI_p$. To check whether *P* is a subset of some existing maximal frequent itemsets, we only need to check them against $LMFI_p$, which takes O($LMFI_p$) cost. If $LMFI_p$ is empty, then *P* will be our new maximal pattern, otherwise it is subset of $LMFI_p$.

In the following sections, we will use *P* as a node in search space and $LMFI_p$ as its local maximal frequent patterns. We will also use $LMFI_{p+1}$ which represents the *LMFI* of child node *P+1*.

## 3. Related Work

Zaki et al in [6] showed the need and importance of LMFI over MFI for the first time, since then almost all MFI algorithms have used LMFI. They introduced the concept of progressive focusing to narrow the search to only the most relevant maximal itemset, making superset checking more powerful. The main idea is to progressively narrow down the maximal itemsets of interest as recursive calls are made. In other words they construct for each invocation an $LMFI_{p+1}$, which contains only relevant maximal patterns. In this way instead of checking superset in *list (MFI)*, we can check it in $LMFI_p$ set. In [6] they also showed the effectiveness of progressive focusing through detailed experiments on different sparse and dense datasets.

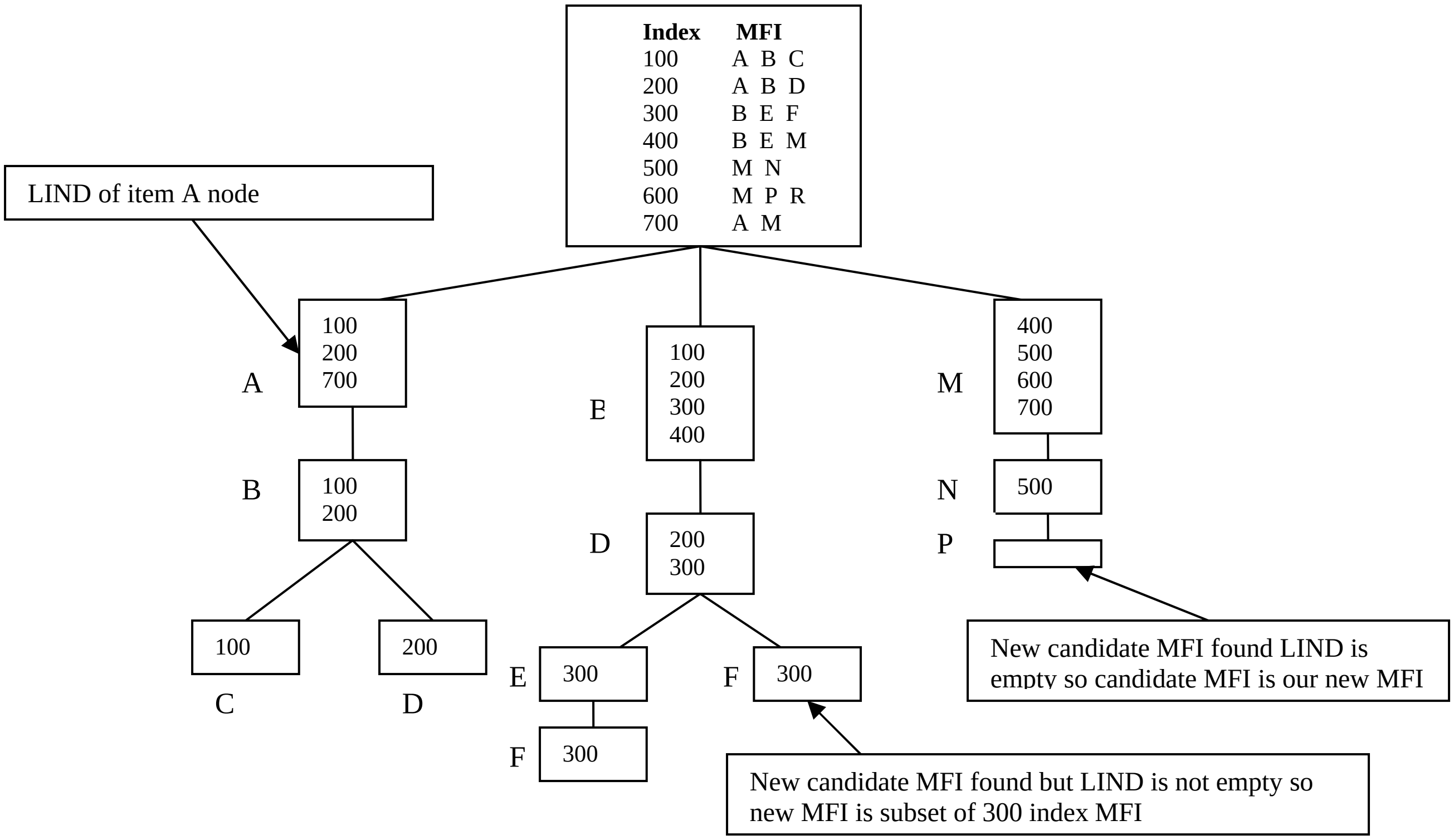

**Figure 1. FastLMFI propagation and FastMFI superset checking example**

## 4. FastLMFI: Local Maximal Patterns Propagation and Maximal Patterns Superset Checking

In this section we explain the LMFI propagation and MFI superset checking using indexing approach. From implementation point of view progressive focusing $LMFI_p$ (where $P$ is any node) can be constructed either from its parent $LMFI_p$ or sibling of $P$. With progressive focusing, construction of child *LMFIs* takes two steps [9]. First, project them in parent $LMFI_{p+1}$. Second, pushing and placing them in top or bottom of *list (MFI)* for constructing $LMFI_{p+1} = LMFI_{p \cup \{i\}}$, where $i$ is tail item of node $P$.

We here list up the some advantages of our indexing approach.

1. Creating child $LMFI_{p+1}$ in one step, rather than into two steps. By using our indexing approach, we can completely eliminate second step. It may be noted the second step is more costly (removing and adding pointers) than first step.

2. Optimizing first step by an efficient implementation (section 5).

### 4.1. Local Maximal Patterns Propagation

With indexing approach we propagate a local index list (LIND) $LIND_{p+1}$ for each tail frequent itemset $FI_{p+1}$, which contains the indexes (positions) of local maximal patterns in *list (MFI).* For example in Figure 1, node A contains the indexes of those local maximal patterns where A appears as a prefix. Child $LIND_{p+1}$ of node $P$ can be constructed by traversing indexes of parent $LIND_p$ and placing them into child $LIND_{p+1}$, which can be done in one step. Line 1 to 2 in Figure 2 shows the creation of $LIND_{p+1} = LIND_{p \cup \{i\}}$ in one step, where line 2 in Figure 2 at same time traverse indexes of parent $LIND_{p \cup \{i\}}$ and create child $LIND_{p+1} = LIND_{p \cup \{i\}}$ indexes.

***Lemma 1:*** Let $P$ be the node of search space and let $LIND_p$ contains its local maximal patterns. Its tail items $LIND_{p+1}$ can be constructed from local maximal patterns indexes of $P$.

***Proof:*** We know that all tail items are $tail\ \{i\} \subseteq P$, and $LIND_p$ contains all those maximal patterns indexes, where $P$ is appear as a prefix. So tail item $LIND_{p+1}$ can be constructed directly from indexes of $LIND_p$, because $LIND_{p+1} \subseteq LIND_p$.

Note that $LIND_p$ of itemset $P$ contains exactly same number of local maximal patterns as progressive focusing $LMFI_p$. Only difference between the two techniques is that, our approach propagate an index list $LIND_{p+1}$ to child nodes, where as progressive focusing pushes and places them in top or bottom of *list (MFI)*.

### 4.2. Incrementing Parent Local Indexes

Note that node $LIND_p$ contains exactly those indexes of maximal patterns which are known to the parent of $LIND_p$. In other words $LIND_p$ does not contain those maximal patterns indexes which are mined or found in sub tree of $P$. To update those indexes found in sub tree of P, we must add all new indexes of $LIND_{p+1}$ into $LIND_p$. Procedure *IncrementSubtreeIndexes (parent LIND, child LIND).* Figure 3 shows the steps of incrementing parent indexes from its child nodes.

***Procedure propagateLIND ($LIND_p$, P)***

```
1  for each tail item of node P i ε tail (P)
2    for each index of LIND_p
3      LIND_{p+1} = LIND_{p ∪ {i}}
4      propagateLIND(LIND_{p+1}, P ∪ {i} )
5      incrementSubtreeIndexes(LIND_p, LIND_{p+1})
```

**Figure 2. Pseudo code of FastLMFI propagation**

***Procedure IncrementSubtreeIndexes ($LIND_p$, $LIND_{p+1}$)***

```
for each l index of LIND_{p+1} not in LIND_p
    LIND_p = LIND_p + l
```

**Figure 3. Pseudo code of incrementing parent local indexes**

### 4.3. Maximal Patterns Superset Checking

If any node $P$ finds a candidate maximal pattern, and if it contains an empty $LIND_p$, then candidate maximal pattern will be our new mined MFI pattern, otherwise it is subset of $LIND_p$ patterns.

Figure 1 shows the process of propagation of $LIND_{p+1}$ and maximal patterns superset checking. Note that the root node contains all the known maximal patterns, which propagate $LIND_{p+1}$ to its child nodes.

***Example 1:*** Let us take an example of propagation of LIND from itemset A to itemset ABC. First, root node propagate itemset A's local maximal pattern indexes {100,200,700} to its child node A, because itemset A appears as a prefix in all these known maximal patterns. In next recursion, node A propagates local maximal pattern indexes to its child nodes, after comparing against its local maximal pattern indexes. Itemset AB is appears as a prefix in {100,200} of node A's local maximal pattern indexes. Where itemset ABC is appears as a prefix in {100} of node AB's local maximal pattern indexes.

## 5. Efficient Implementation of FastLMFI

### 5.1. Local Maximal Patterns Representation

We choose to use a vertical bitmap for the mined maximal patterns representation. In a vertical bitmap, there is one bit for each maximal pattern. If item i appears in maximal pattern j, then the bit of j of the bitmap of item i is set to one; otherwise the bit is set to zero. Figure 4 shows the vertical bitmap representation of maximal patterns.

**A B C**
**B D E F G**
**E H**
**C E G**
**A C G**
**A C D F H**
**F I**

| Transaction | A | B | C | D | E | F | G | H | I |
|---|---|---|---|---|---|---|---|---|---|
| 1 | 1 | 1 | 1 | 0 | 0 | 0 | 0 | 0 | 0 |
| 2 | 0 | 1 | 0 | 1 | 1 | 1 | 1 | 0 | 0 |
| 3 | 0 | 0 | 0 | 0 | 1 | 0 | 0 | 1 | 0 |
| 4 | 0 | 0 | 1 | 0 | 1 | 0 | 1 | 0 | 0 |
| 5 | 1 | 0 | 1 | 0 | 0 | 0 | 1 | 0 | 0 |
| 6 | 1 | 0 | 1 | 1 | 0 | 1 | 0 | 1 | 0 |
| 7 | 0 | 0 | 0 | 0 | 0 | 1 | 0 | 0 | 1 |

Maximal patterns

Vertical bitmap representation

**Figure 4. A sample of maximal patterns with vertical bitmaps representation**

Note that each index of $LIND_p$ points to some position in $P = \{0 \cup 1 \cup 2 \cup \ldots n\}$ bitmap. $P$ child $LIND_{p+1}$ can be constructed by taking *AND* of $LIND_p$ bitmap, with tail item $X$ bitmap.

**bitmap** ($LIND_{p+1}$) = **bitmap** ($LIND_p$) AND **bitmap**($X$)

There are two ways of representing maximal patterns for each index of $LIND_p$. First, way is that each index of $LIND_p$ points to exactly one maximal pattern. Second, way can be each index of $LIND_p$ points to 32 maximal patterns of whole 32-bit integer range. The second approach was used for fast frequency counting in [4] and they show that it is better than single bit approach with a factor of 1/32. We also observed through experiments that second approach is more efficient than first approach for local maximal patterns propagation. Figure 5 compares the 32-maximal patterns per index with single maximal pattern per index, on retail dataset with different support thresholds.

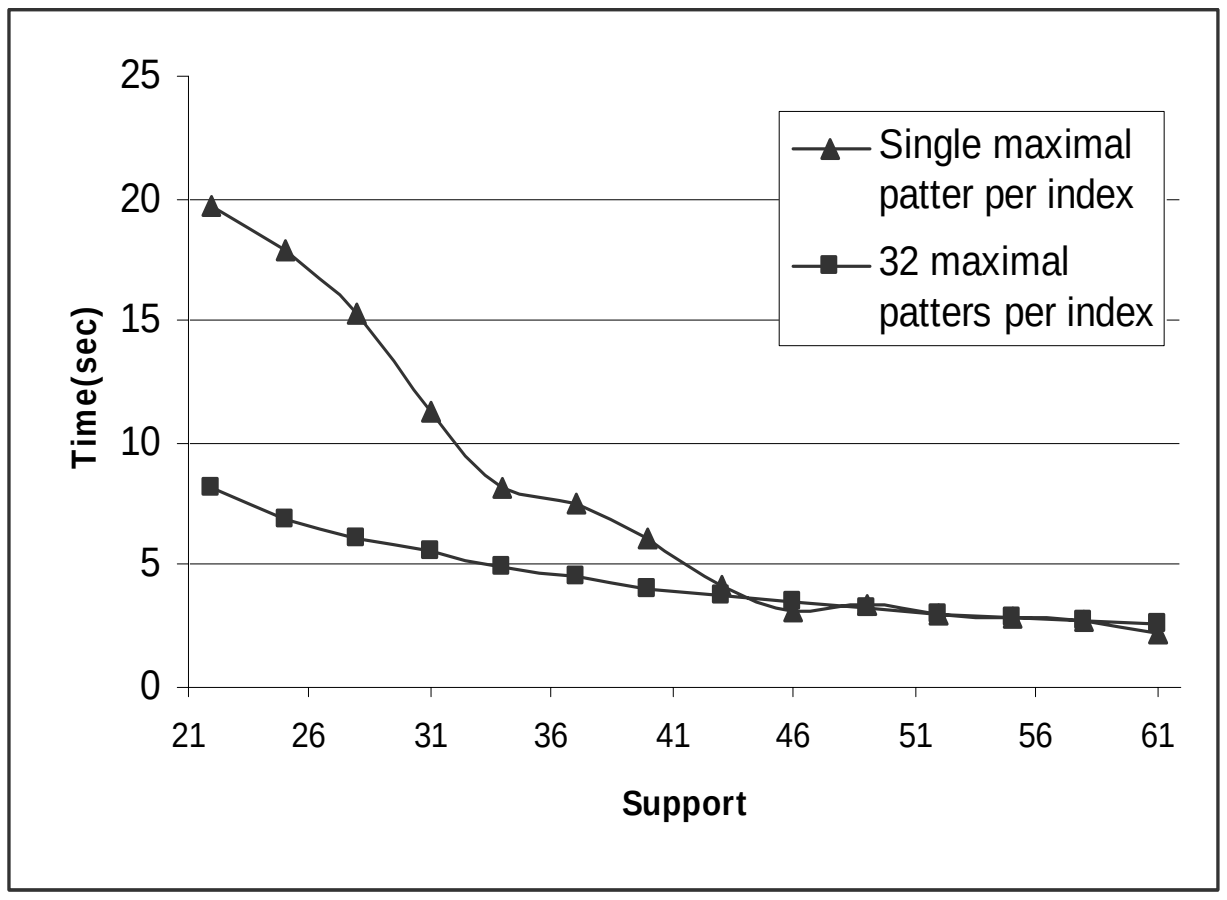

**Figure 5. Indexing with single maximal pattern versus indexing with 32-maximal patterns**

### 5.2. Memory Optimization

As explained earlier each recursion of MFI algorithm constructs and propagates $LIND_{p+1}$ to its child nodes. One way of construction of child $LIND_{p+1}$ is to declare a new memory and then propagate to child nodes. Obviously this technique is not space efficient. A better approach is as follows. We know that with Depth First Search (DFS) a single branch is explored at any time. Before starting the algorithm we create a large memory (equal to all known maximal patterns) for each level, which is equal to the maximal branch length. Next time each level of DFS tree can share this memory, and does not need to create any extra memory at each recursion level.

## 6. Implementation and Results

The code of FastLMFI is written in Visual C++ 6.0 with HybridMiner MFI algorithm [2]. Experiments have been conducted on the Celeron (1.00 GHz) processor with main memory of size 160 MB.

In this section we describe the performance of FastLMFI versus progressive focusing on the benchmark datasets downloaded from http://fimi.cs.helsinki.fi/fimi03/datasets.html. The main features of these datasets are listed in Table 1.

**Table 1. Main features of datasets**

| Dataset | Items | Average Length | Records |
|---|---|---|---|
| T10I4D100K | 1000 | 10 | 100,000 |
| T40I10D100K | 1000 | 40 | 100,000 |
| Kosarak | 20,753 | 8.1 | 66,283 |
| Retail | 16,469 | 10.3 | 88,162 |
| Chess | 75 | 35 | 3,196 |
| Mushroom | 119 | 23 | 8,124 |

**Mushroom:** A dataset with information about various mushroom species.
**Chess:** Contains game data of chess.
**T10I4D100K, T40I10D100K:** Synthetic datasets.
**Kosarak:** Contain click-stream data of a Hungarian on-line news portal.
**Retail:** Retail market basket data from an anonymous Belgian retail store.

The performance measure is the execution time of the progressive focusing [6] versus FastMFI on different support threshold using Table 1 datasets. For clarity we omit three subset pruning techniques from our results. We refer [4] to readers for detailed of subset pruning (PEP, FHUT, HUTMFI) techniques. A short definition of all sub space pruning techniques is given below.

***FHUT:*** Let *P* be the node of search space with head *X* and tail *Y*. If *{XUY}* is a maximal frequent itemset, then all subsets of tail *Y* combined with head *X* are also frequent but not maximal, and can be pruned away.

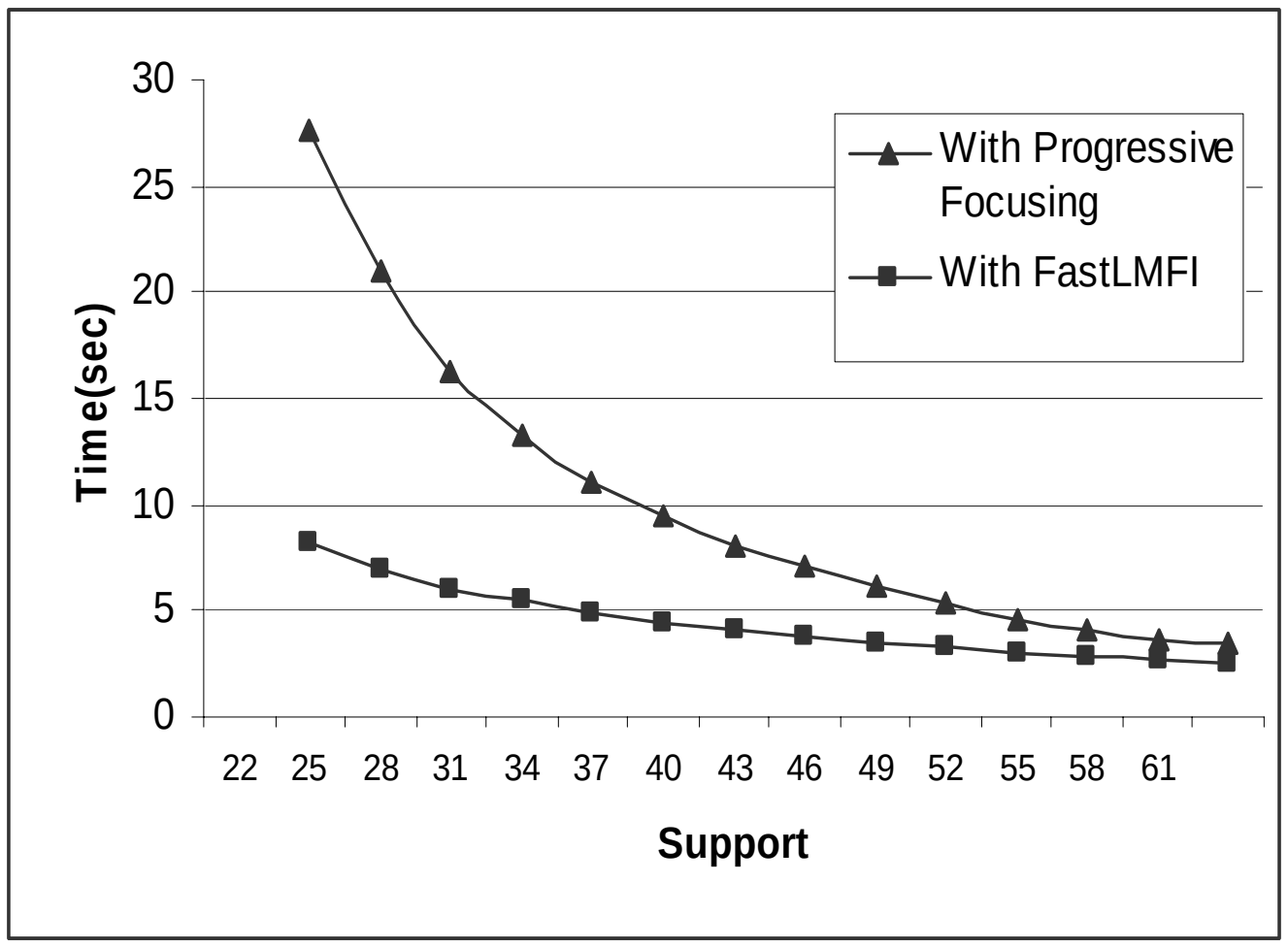

**Figure 6. Retail dataset**

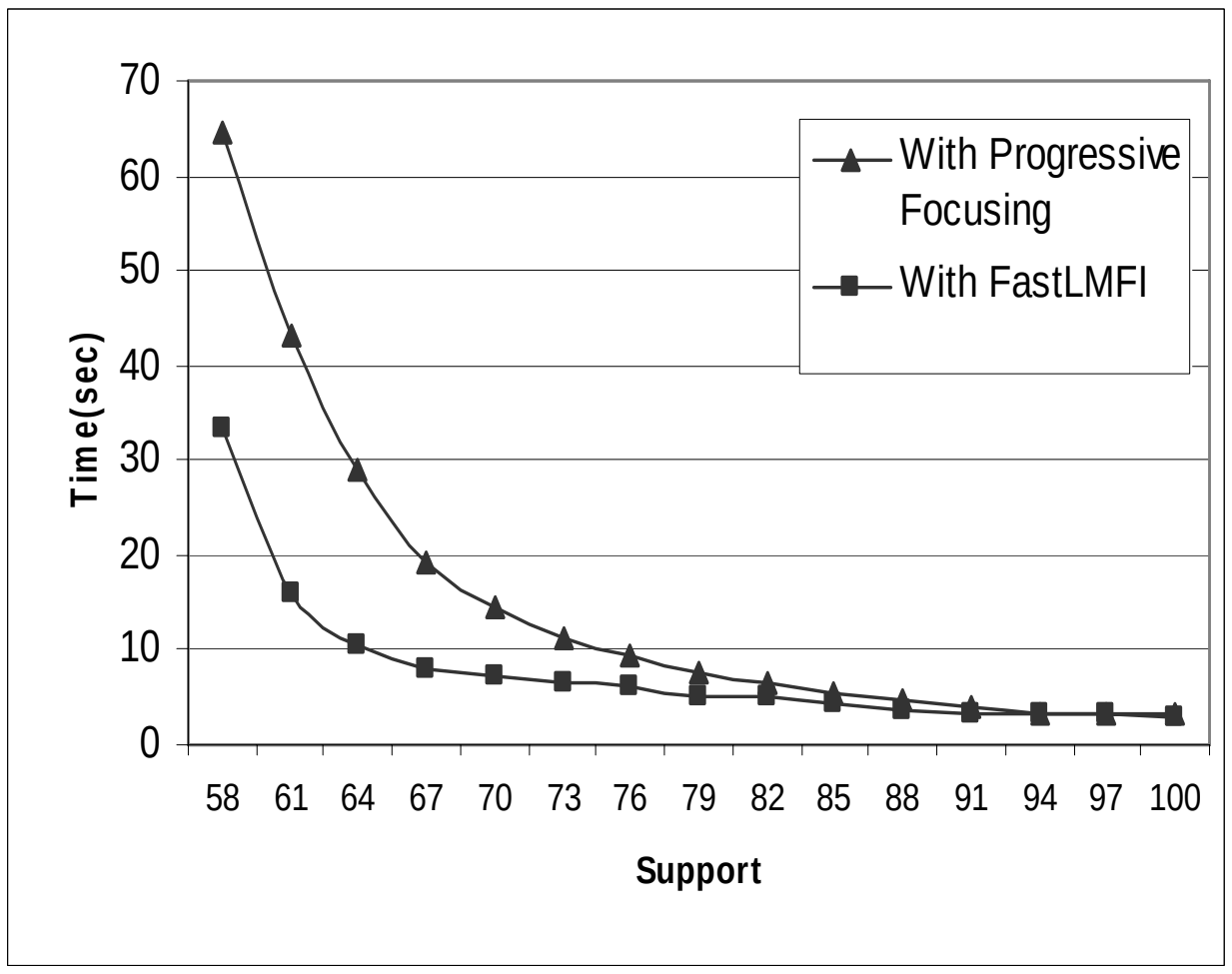

**Figure 7. Kosarak dataset**

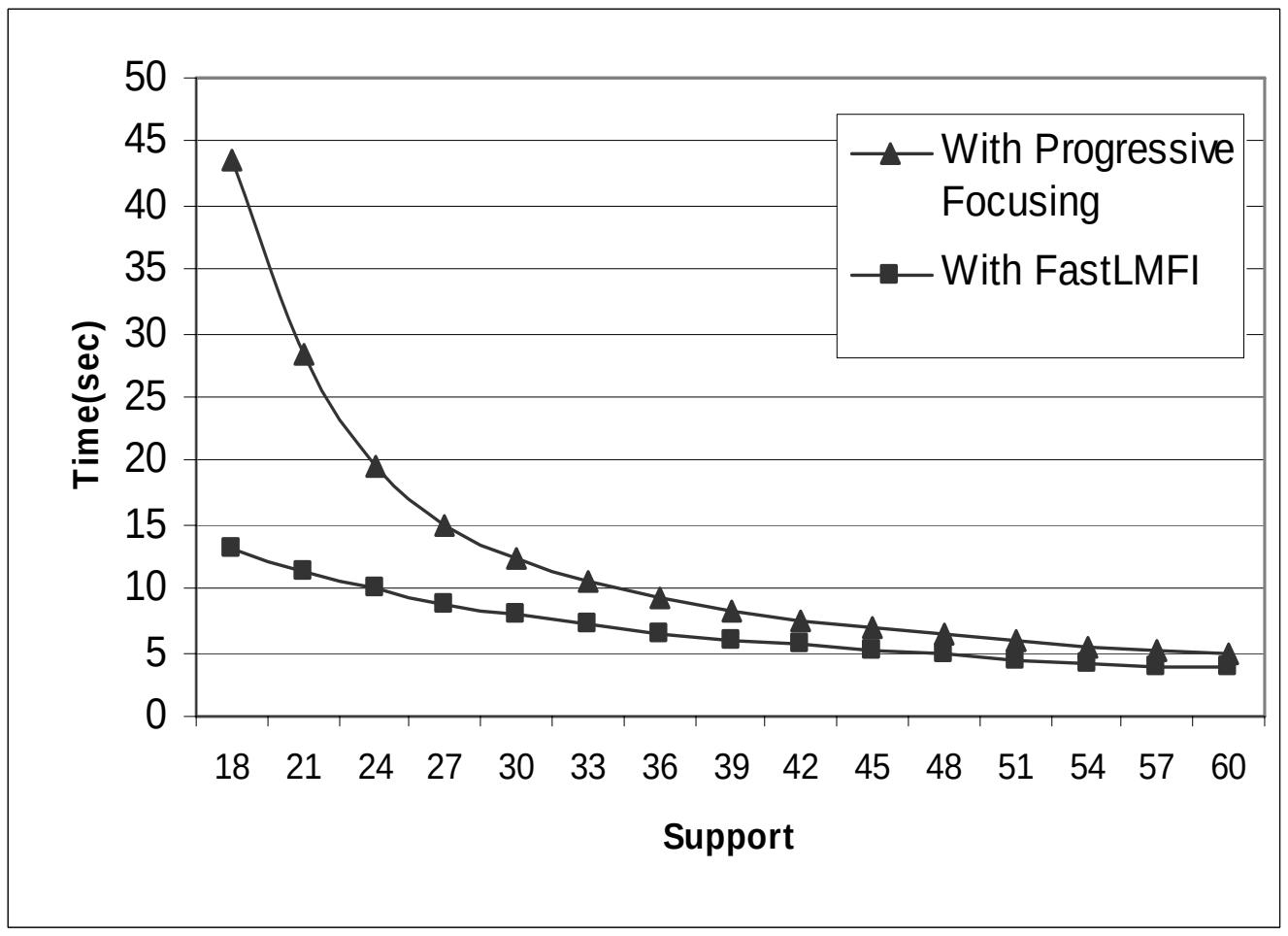

**Figure 8. T10I4D100K dataset**

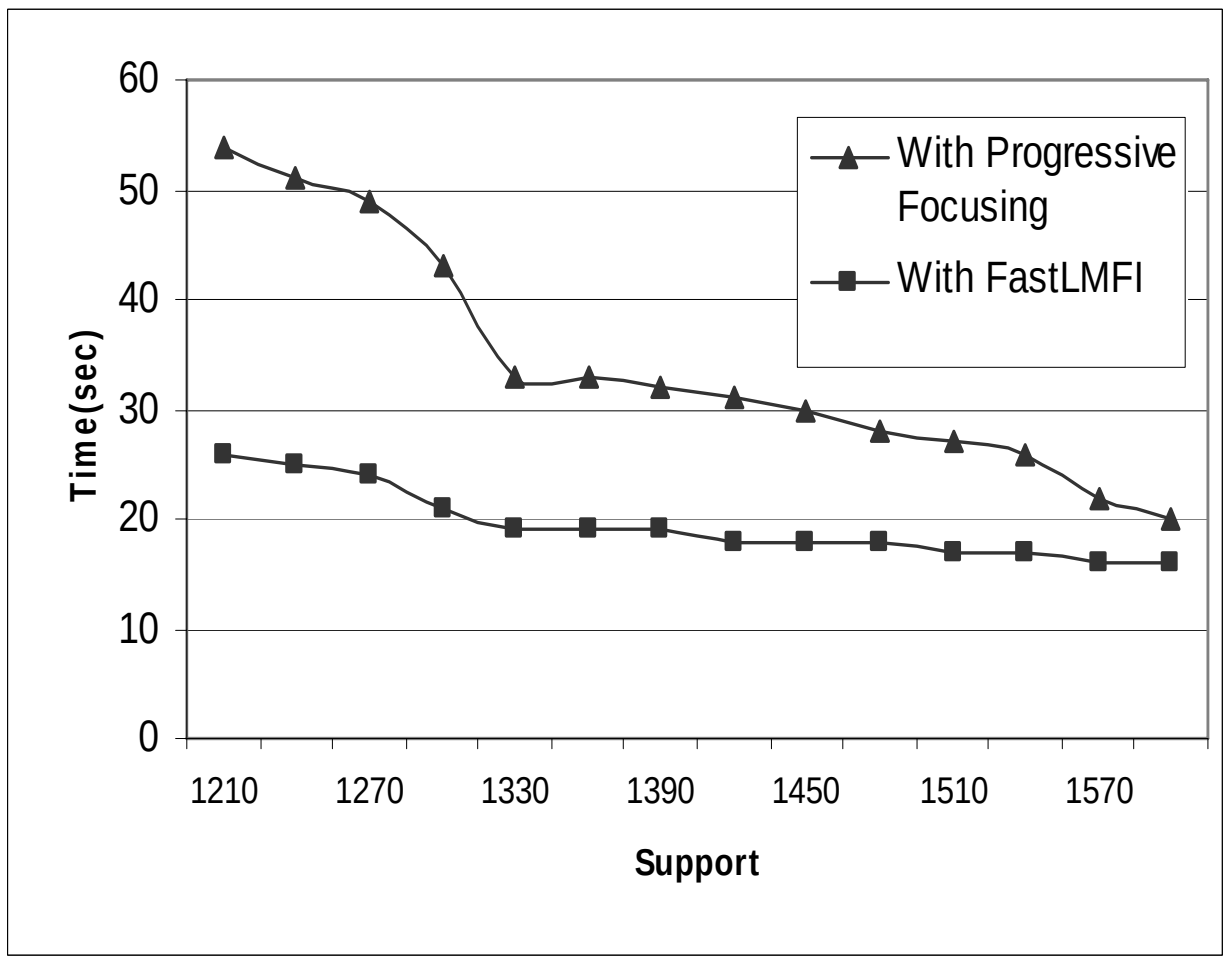

**Figure 9. T40I10D100K dataset**

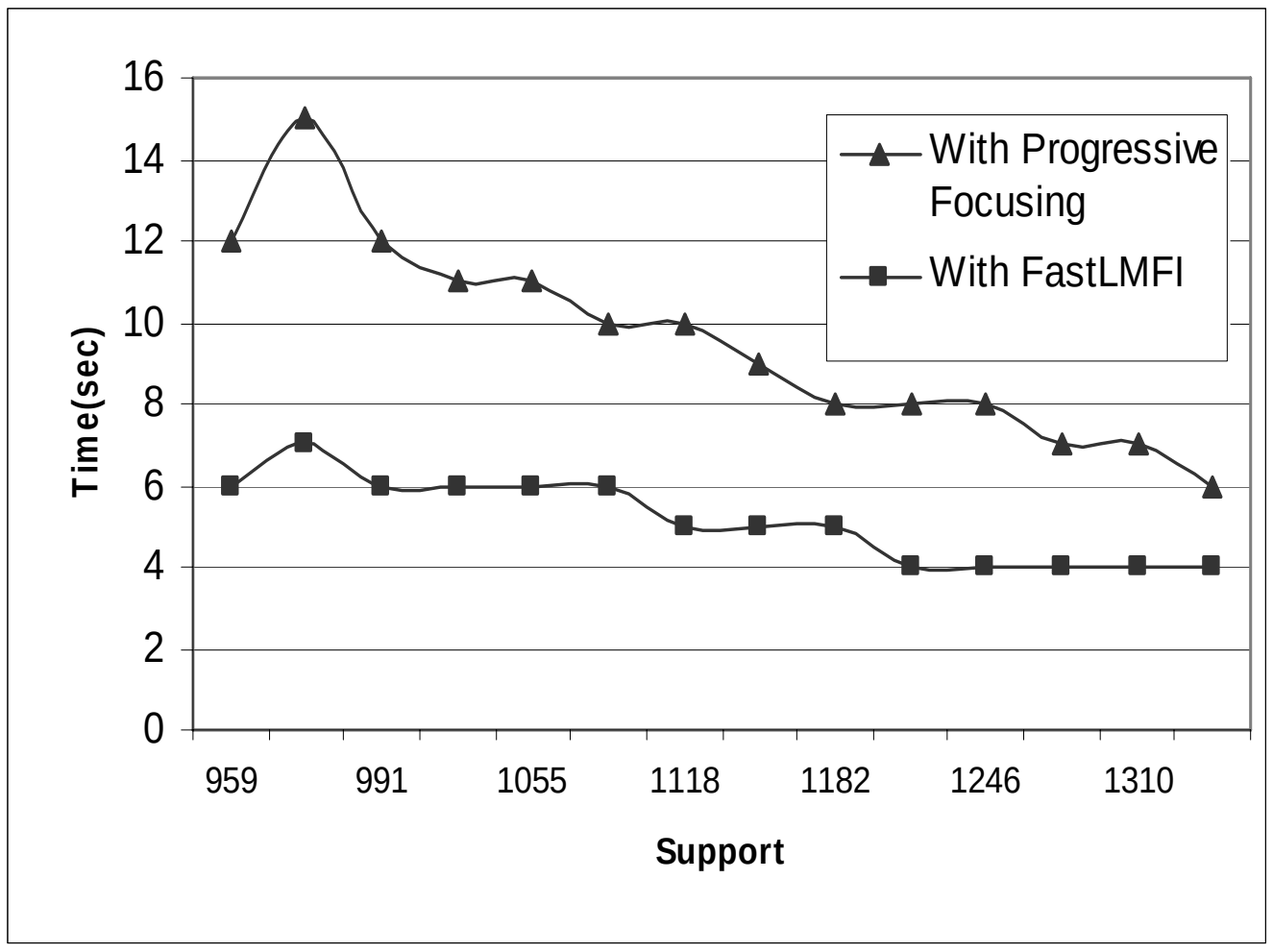

**Figure 10. Chess dataset**

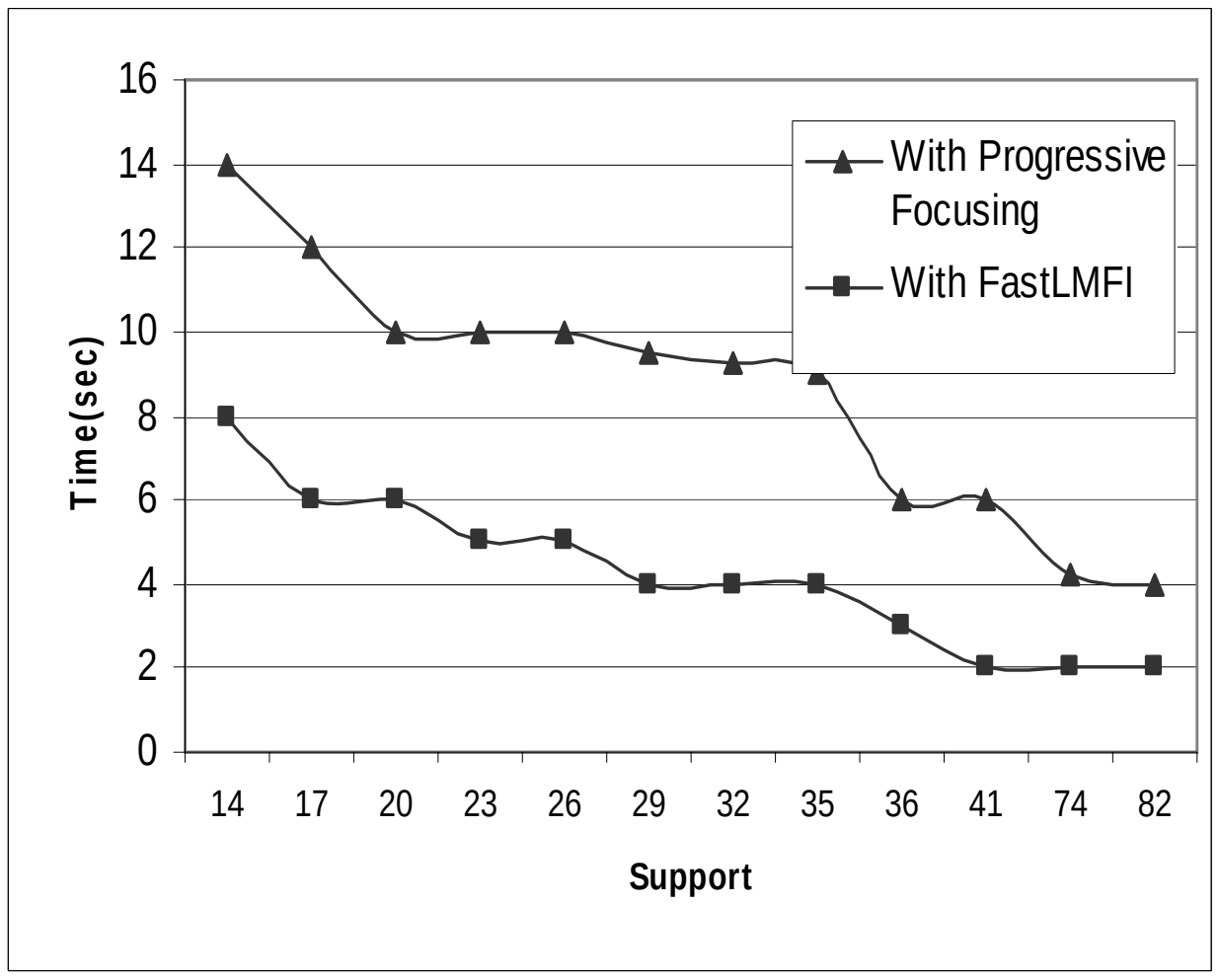

**Figure 11. Mushroom dataset**

***HUTMFI:*** Let *P* be the node of search space with head *X* and tail *Y*. If tail *Y* is the subset of any known maximal frequent itemset, then whole sub tree and sibling is pruned away.

***PEP:*** Let *P* be the node with head *X* and tail *Y*. If *Y* element *S* has same support as head *X*, then *S* is moved from tail to head. We know that *transactions(X)* $\subseteq$ *transactions (S)*.

Figures from 6 to 11 show the performance curve of the two techniques. As we can see, the FastLMFI is better than progressive focusing on sparse datasets as well as dense datasets. The performance improvements of FastLMFI over progressive focusing are significant at reasonably low support thresholds.

# 7. FastLMFI Integration with Mafia

Mafia [4] used progressive focusing for maximal itemset superset checking. As reported in [5], Mafia is considered to be most efficient MFI algorithm for small dense (chess, mushroom) datasets. In this section explain the integration results of FastLMFI superset checking with Mafia. Our experimental results on small dense (chess, mushroom) datasets show that, we can further improve the performance of any MFI algorithm by performing superset checking using our approach.

Figure 13 and Figure 14 show the computational results of Mafia superset checking by using FastLMFI comparing with Mafia, afopt_max [9] and FP (zhu)-max [7] which received top scores in FIMI03 and FIMI04 [5]. Our results show that Mafia-FastLMFI not only outperforms Mafia-Progressive Focusing algorithm, but its results are also better than current two best afopt_max and FP (zhu)-max (as reported in [5]) algorithms. Mafia, afopt_max and FP (zhu)-max implementations are available at http://fimi.cs.helsinki.fi/fimi03/implementations.html.

## 7.1. Mafia

Mafia [4] proposed parent equivalence pruning (PEP) and differentiates superset pruning into two classes FHUT and HUTMFI. For a given node ***X:aY***, the idea of PEP is that if ***sup(X)=sup(Xa)***, i.e. every transaction containing ***X*** also contains the item ***a***, then the node can simply be replaced by ***Xa:Y***. The FHUT uses leftmost tree to prune its sister, i.e., if the entire tree with root ***Xa:Y*** is frequent, then we do not need to explore the sisters of the node ***Xa:Y***. The HUTMFI uses to use the known MFI set to prune a node, i.e., if itemset of ***XaY*** is subsumed by some itemset in the MFI set, the node ***Xa:Y*** can be pruned. Mafia also uses dynamic reordering to reduce the search space. The results show that PEP has the biggest effect of the above pruning methods (PEP, FHUT, and HUTMFI). The pseudo code of FastLMFI integration with Mafia is shown in Figure 12.

**Procedure Mafia-FastLMFI ($LIND_p$, P, Boolean IsHUT)**

```
variable HUT = P(head) ∪ P(tail)
if HUT is in LIND_p
    stop generation of child node and return

reorder by increasing support, use PEP
to trim the tail
for each item i in P(trimmed_tail)
 {
      isHUT = whether i is the first item in the tail
      newNode = P ∪ i
      LIND_{p+1} = LIND_p ∪{i}
      Mafia-FastLMFI (newNode, LIND_{p+1}, isHUT)
 }

if ( isHUT and all extensions are frequent )
      stop search and go back up subtree
if ( P is a leap and LIND_p is empty )
      Add P(head) to list (MFI)
```

**Figure 12. Pseudo code of FastLMFI (superset checking) integration with Mafia**

## 7.2. Computational Results

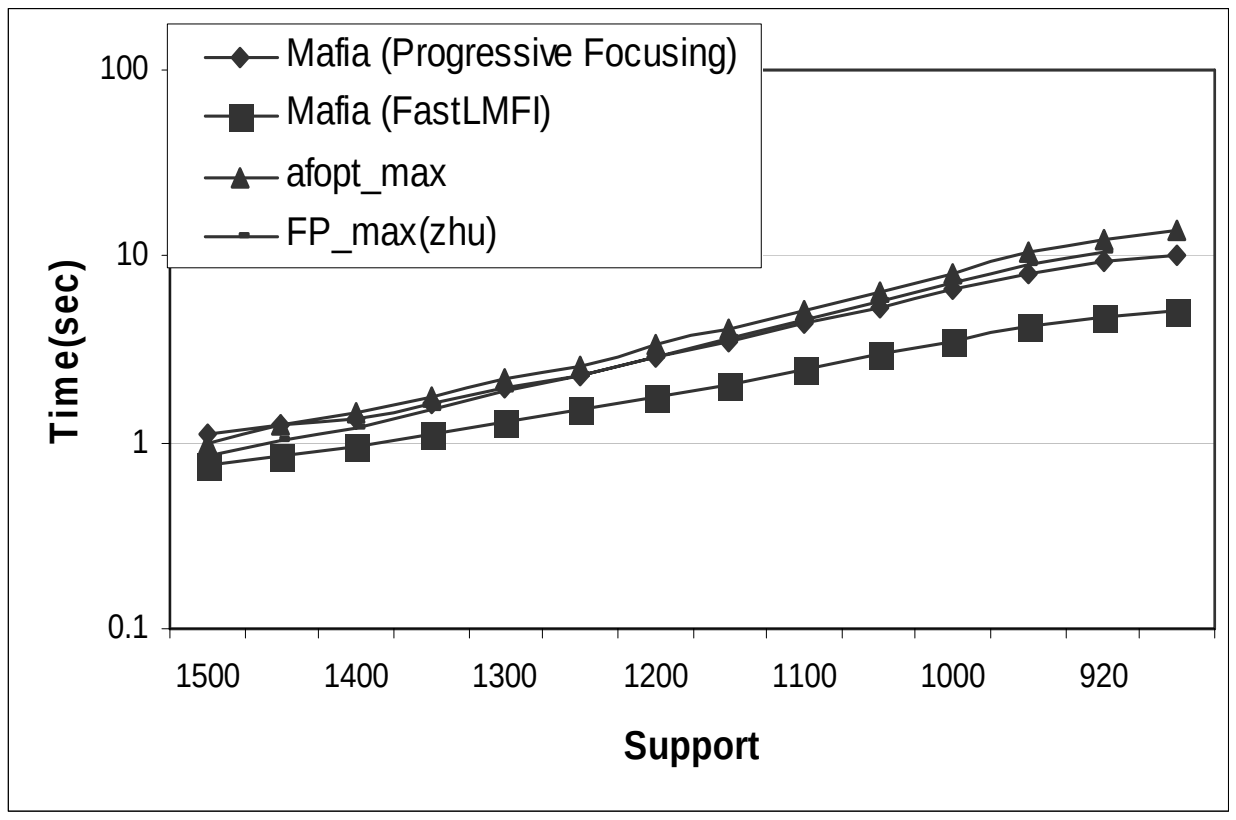

**Figure 13. Performance result on Chess dataset**

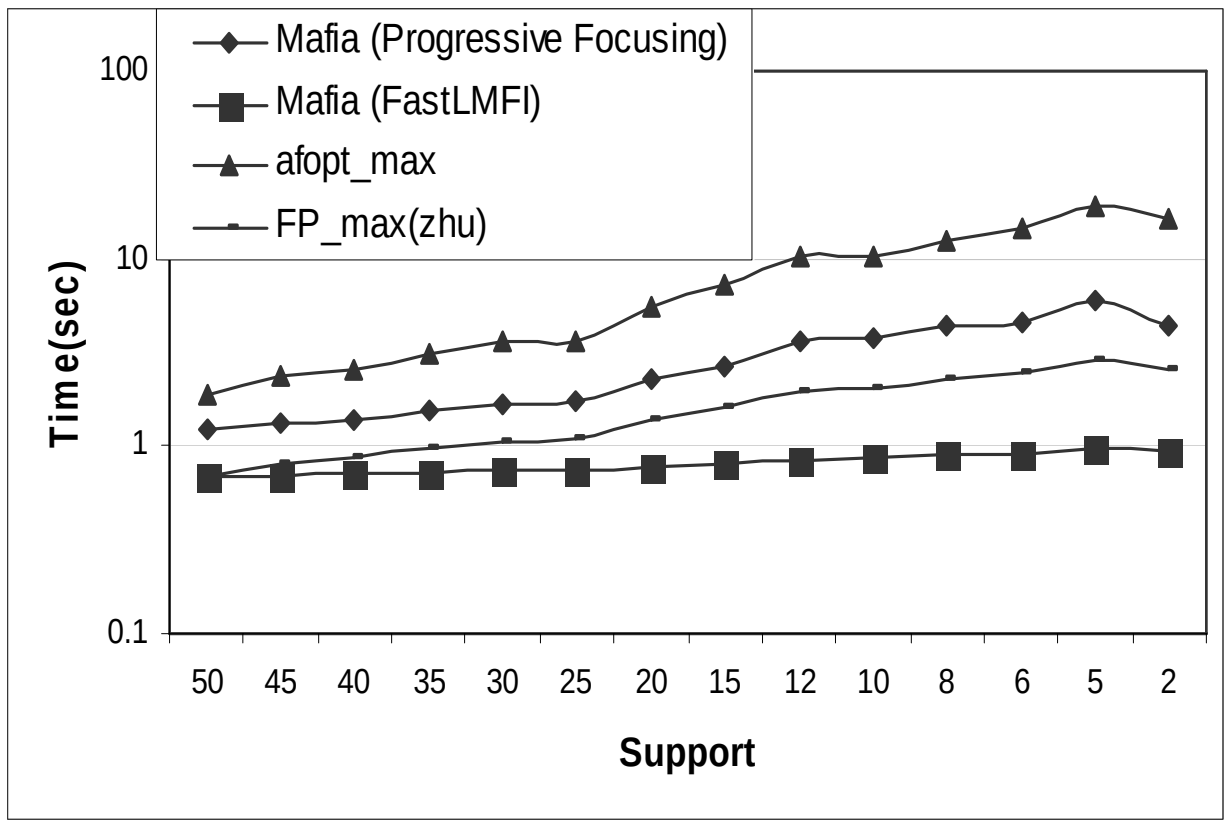

**Figure 14. Performance result on Mushroom dataset**

## 8. Conclusion

Maximal itemset superset check is considered to be an important factor in overall MFI mining. As we have seen from different experimental results its cost is almost half of the total MFI mining cost. In this paper we have present a new approach ***FastLMFI*** for local maximal patterns propagation and maximal patterns superset checking. Different components of ***FastLMFI*** show that construction of child local maximal patterns *LINDs* in one step rather than two steps, and representing 32 maximal patterns per index of *LIND* is smart and fast approach than previous well known ***progressive focusing*** approach.

## Acknowledgement

We are very thankful to Prof. Bart Goethals for his efforts on maintaining FIMI repository. We are also thankful to all authors that gave their implementations and datasets in FIMI repository.